\documentclass[aps,prl,twocolumn,superscriptaddress]{revtex4-1}
\usepackage{stackengine,graphicx} 
\usepackage{epsfig} 
\usepackage{dcolumn} 
\usepackage{bm} 
\usepackage{xcolor}
\usepackage{xspace}
\usepackage[normalem]{ulem}

\graphicspath{{./}{Figures//}}
\newcounter{saveenumi}

\newcommand{\be}{\begin{enumerate}}
\newcommand{\ee}{\end{enumerate}}
\newcommand{\ignore}[1]{}

\definecolor{robgreen}{rgb}{0.333,0.62,0.18}
\definecolor{linebrown}{HTML}{bd6e00}
\definecolor{lineblue}{HTML}{0000ff}
\definecolor{linegreen}{HTML}{217200}
\definecolor{linered}{HTML}{d20d0d}

\def\cis{CuIr$_{2}$S$_{4}$}

\def\mto{MgTi$_{2}$O$_{4}$}
\def\ntso{NaTiSi$_{2}$O$_{6}$}
\def\ltso{LiTiSi$_{2}$O$_{6}$}
\def\fd3m{Fd$\overline 3$m}
\def\p1bar{P$\overline 1$}
\def\c2c{C$2/c$}
\def\i41amd{I4$_{1}$/amd}
\def\t2g{$t_{2g}$}

\def\tio6{TiO$_6$}
\def\plaquette{TiO$_2$}
\def\rangetxpd{90~K$\le T\le$ 300~K}

\def\rrange{1~\AA$ \le r\le$ 15~\AA}
\def\rrangee{15~\AA$ \le r\le$ 30~\AA}

\newcommand{\ins}[1]{\textcolor{brown}{[ins:#1]}}
\renewcommand{\sout}[1]{}
\renewcommand{\ins}[1]{#1}
\begin{document}
%
%
\begin{abstract}
The local structure of \ntso\ is examined across its Ti-dimerization orbital-assisted Peierls transition at 210~K.
An atomic pair distribution function approach evidences local symmetry breaking preexisting far above the transition.
The analysis unravels that on warming the dimers evolve into a short range orbital degeneracy lifted (ODL) state of dual orbital character, persisting up to at least 490~K.
The ODL state is correlated over the length scale spanning $\sim$6 sites of the Ti zigzag chains.
Results imply that the ODL phenomenology extends to strongly correlated electron systems.
\end{abstract}

\title{Dual Orbital Degeneracy Lifting in a Strongly Correlated Electron System}

\author{R.~J.~Koch}\email[]{rkoch@bnl.gov}
\affiliation{Condensed Matter Physics and Materials Science Division, Brookhaven National Laboratory, Upton, NY 11973, USA}
\author{R.~Sinclair}
\affiliation{Department of Physics and Astronomy, University of Tennessee, Knoxville, TN 37996, USA}
\author{M.~T.~McDonnell}\altaffiliation[Present address: ]{Computer Science and Mathematics Division, Oak Ridge National Laboratory, Oak Ridge, TN 37831, USA}
\affiliation{Neutron Scattering Division, Oak Ridge National Laboratory, Oak Ridge, TN 37831, USA}
\author{R.~Yu}\altaffiliation[Present address: ]{Institute of Physics, Chinese Academy of Science, Beijing 100190, People’s Republic of China}
\affiliation{Condensed Matter Physics and Materials Science Division, Brookhaven National Laboratory, Upton, NY 11973, USA}
\author{M.~Abeykoon}
\affiliation{Photon Sciences Division, Brookhaven National Laboratory, Upton, NY 11973, USA}
\author{M.~G.~Tucker}
\affiliation{Neutron Scattering Division, Oak Ridge National Laboratory, Oak Ridge, TN 37831, USA}
\author{A.~M.~Tsvelik}
\affiliation{Condensed Matter Physics and Materials Science Division, Brookhaven National Laboratory, Upton, NY 11973, USA}
\author{S.~J.~L.~Billinge}
\affiliation{Condensed Matter Physics and Materials Science Division, Brookhaven National Laboratory, Upton, NY 11973, USA}
\affiliation{Department of Applied Physics and Applied Mathematics, Columbia University, New York, NY~10027, USA}
\author{H.~D.~Zhou}
\affiliation{Department of Physics and Astronomy, University of Tennessee, Knoxville, TN 37996, USA}
\author{W.-G.~Yin}
\affiliation{Condensed Matter Physics and Materials Science Division, Brookhaven National Laboratory, Upton, NY 11973, USA}
\author{E.~S.~Bozin}\email[]{bozin@bnl.gov}
\affiliation{Condensed Matter Physics and Materials Science Division, Brookhaven National Laboratory, Upton, NY 11973, USA}

\date{\today}
\maketitle

The rich physics associated with the emergence of technologically relevant quantum orders in materials~\cite{tokura_emergent_2017} stems from complex interaction of electronic charge, spin, and orbitals, and their coupling to the host lattice~\cite{editorial_rise_2016,keimer_quantum_2015}.
In transition metal systems with partial filling of $d$-manifolds novel properties often engage the orbital sector~\cite{tokura_orbital_2000}.
Systems exhibiting orbital degeneracy and/or electronic frustration imposed by their lattice topology are of particular interest, as orbitals couple both to the spin, via electronic interactions, and to the lattice, via Jahn-Teller type mechanisms~\cite{streltsov_orbital_2017,vojta_frustration_2018}.
The removal of this orbital degeneracy and the subsequent relief of frustration then impact symmetry lowering and material properties.
\ins{The e}\sout{E}lectronic complexity of the low temperature ordered symmetry-broken states has been thoroughly studied in systems displaying diverse emergent behaviors such as frustrated magnetism~\cite{zorko_frustration-induced_2014,glasbrenner_effect_2015}, colossal magnetoresistivity~\cite{savitzky_bending_2017}, charge and orbital order~\cite{radaelli_formation_2002,achkar_orbital_2016}, metal-insulator transition~\cite{aetukuri_control_2013,tian_field-induced_2016,liang_orthogonal_2017}, pseudogap~\cite{borisenko_pseudogap_2008,ren_direct_2020} and high temperature superconductivity~\cite{wang_nematicity_2015,sprau_discovery_2017}.
Their understanding employs Fermi surface nesting~\cite{johannes_fermi_2008,terashima_fermi_2009}, Peierls~\cite{lee_fluctuation_1973,bhobe_electronic_2015}, and band Jahn-Teller mechanisms~\cite{kanamori_crystal_1960,huang_jahn-teller_2017}, among others.

In systems\sout{ with anticipated} \ins{where }orbital degeneracies\ins{ are anticipated,} crystallographic symmetry lowering at the temperature driven structural phase transitions is often assumed to imply
{\it simultaneous} orbital degeneracy lifting (ODL) by engaging some cooperative mechanism~\cite{oles_spin-orbital_2006,streltsov_orbital_2017,vojta_frustration_2018}.
Consequently, seemingly mundane high temperature regimes possessing high crystallographic symmetry remain much less explored.
In contrast to this concept, recent utilization of probes sensitive to local symmetry have qualified the ODL as a {\it local electronic effect} existing at temperature well above~\cite{bozin_local_2019,long;arxiv20} the global symmetry breaking transitions.
Focusing on the \cis~\cite{bozin_local_2019} and \mto~\cite{long;arxiv20} spinel systems in the weak electron coupling limit, the studies unmasked a highly localized ODL state at high temperature involving two transition metal ions, which serves as a precursor to an orbitally driven metal-insulator transition~\cite{khomskii_orbitally_2005}.
Albeit discontinuously connected to the ground state, the ODL in these spinels proves to be a prerequisite state for charge \&\ orbital order and spin singlet dimerization observed at low temperature~\cite{radaelli_formation_2002,PhysRevLett.92.056402,bozin_detailed_2011}, thus enabling the transition.
The ODL state further rationalizes the apparent order of magnitude discrepancy between the energy scales of the observed phenomena ($\sim$10s-100s of meV corresponding to $\sim$100s-1000s K) and the symmetry lowering transition temperatures (typically $\sim$10s-100s K)~\cite{khoms;b;tmc14,bozin_local_2019}.

\begin{figure}[tb]
\includegraphics[width=0.485\textwidth]{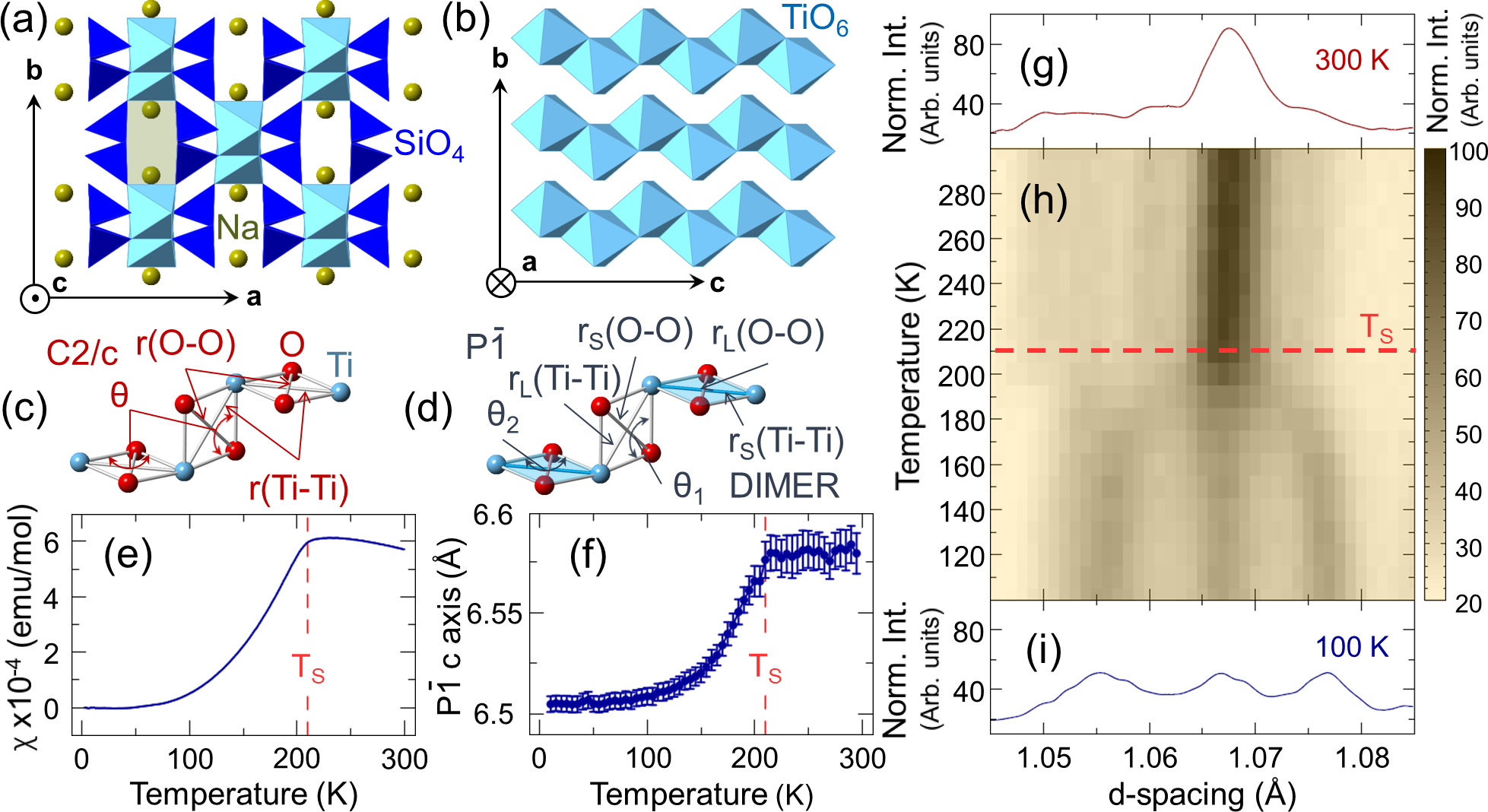}
\caption{\label{fig:propfig} Properties of \ntso: (a) \c2c\ structure; (b) Quasi-1D zigzag \tio6\ chains; (c) Undistorted TiO$_{2}$ plaquettes of the \c2c\ phase featuring uniform Ti-Ti and O-O distances; (d) Distorted TiO$_{2}$ plaquettes of the dimerized \p1bar\ phase with Ti-Ti and O-O distances  bifurcated (S=short, L=long); (e) The Curie law subtracted DC magnetic susceptibility; (f) The c axis parameter from \p1bar\ model fits to the xPDF data; (g)-(i) Temperature evolution of a selected segment of neutron total scattering data. Note: zigzag chains run along $c$ axis in \c2c, and along $a$ axis in \p1bar.}
\end{figure}

The ubiquity and role of the ODL in the emergent quantum orders are yet to be established~\cite{koch;prb19,long;arxiv20}.
\sout{An important aspect}\ins{It is important} to understand \sout{is }whether the ODL state is just a peculiarity of weakly coupled electronic systems in the proximity to a localized\ins{-}\sout{ }to\ins{-}\sout{ }itinerant crossover, or \ins{if }the state could also be realized deep in the Mott insulating regime with strong on-site Coulomb interactions, where the charge fluctuations are suppressed.
Curiously, an opportunity to explore this is offered by the quasi-one dimensional \ntso\ clinopyroxene~\cite{ohashi_crystal_1982}, one of the rock-forming silicate minerals constituting the upper Earth's mantle~\cite{ander;b;ntote12}.
It is a paramagnetic strongly correlated Mott insulator with a $\sim$2~eV gap~\cite{streltsov_comment_2006}, featuring zigzag chains of skew edge-shared \tio6\ (Fig.~\ref{fig:propfig}(a), (b)), with Ti$^{3+}$ in $d^{1}$ ($S = 1/2$) nominally triply degenerate \t2g\ orbital configuration.
A nonmagnetic ground state of \ntso\ with a 53~meV spin gap~\cite{silverstein_direct_2014} establishes on cooling through\ins{ a} 210~K~\cite{isobe_novel_2002} structural transition where orbital ordering stabilizes intrachain Ti-Ti spin singlet dimerization~\cite{streltsov_electronic_2008}.
Once thought to host Haldane $S = 1$ chains~\cite{popovic_sodium_2004,streltsov_comment_2006,popovic_popoviifmmode_2006}, \ntso\ is considered a candidate for quantum liquid with strong orbital fluctuations~\cite{feiguin_quantum_2019}.

By combining neutron and x-ray total scattering based atomic pair distribution function (nPDF and xPDF) approaches~\cite{egami;b;utbp12} we find compelling local structural evidence for a fluctuating ODL state of dual orbital character in \ntso\ at high temperature.
The spatial extent of associated short range structural correlations implies Peierls-like instability at 1/6 filling and \ins{a }relevance of all three Ti \t2g\ orbitals in that regime.
The PDF observations establish that the ODL phenomenology does extend to materials with strong electron correlations, reinforcing the notion of its ubiquity. The results account for a number of high temperature anomalies reported in previous studies of this system~\cite{silverstein_direct_2014,konstantinovic_orbital_2004,rivas-murias_rapidly_2011}, and provide new insight into the understanding of the transition mechanism.

Polycrystalline \ntso\ used in powder diffraction measurements was obtained via a solid state route~\cite{isobe_novel_2002,silverstein_direct_2014,supplementalMaterials} and shows a transition to a nonmagnetic state below T$_{s}=210$~K, Fig.~\ref{fig:propfig}(e).
Total scattering data for PDF analysis were collected over 100~K$\le$T$\le$300~K (neutrons), and over 10~K$\le$T$\le$300~K range and at 490~K (x-rays).
The approach utilizes both Bragg and diffuse scattering, and provides information on the average structure and on the local deviations from it~\cite{egami;b;utbp12}.
Robust crystallographic symmetry change at the transition is evident in the x-ray (Fig.~\ref{fig:propfig}(f)), and neutron data (Figs.~\ref{fig:propfig}(g)-(i)), where there is an abrupt discontinuity in the number, intensity, and strength of observed Bragg peaks in reciprocal space.
Details of data collection and reduction, and PDF analysis protocols used~\cite{egami;b;utbp12} are provided elsewhere~\cite{supplementalMaterials}.

\begin{figure}[tb]
\includegraphics[width=0.485\textwidth]{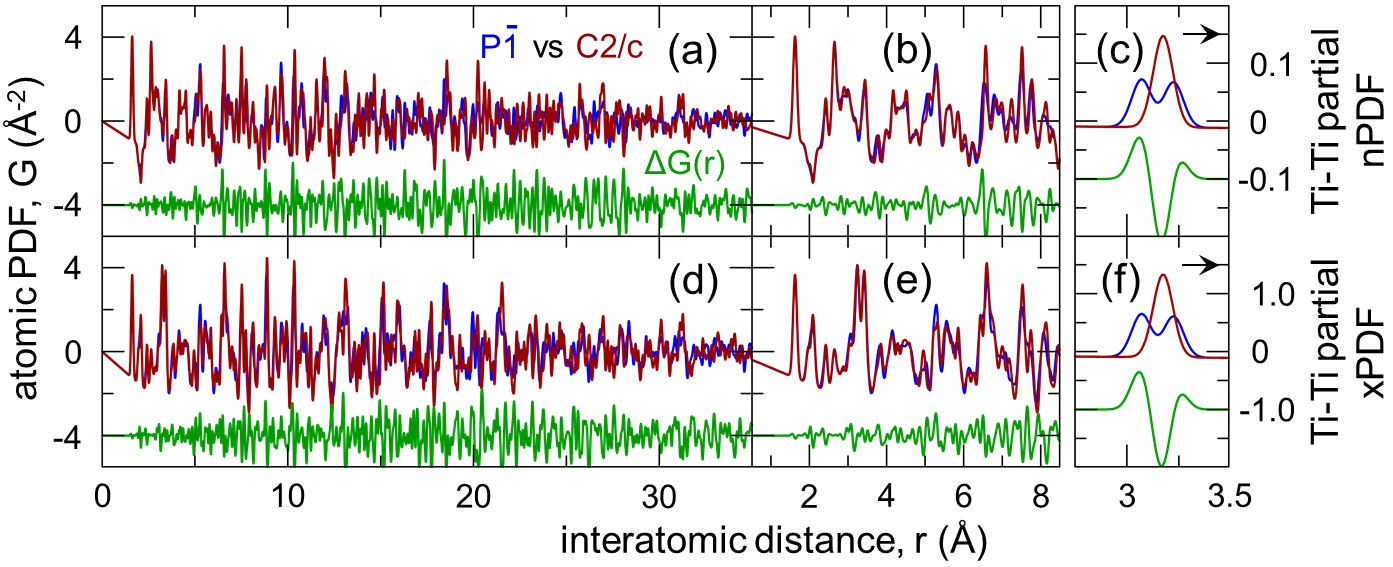}
\caption{\label{fig:simPDFcomp} Comparison of simulated PDFs: Crystallographic \p1bar\ (blue) and \c2c\ (red) models with their differential (green) offset for clarity for neutron probe over a wide (a) and a narrow (b) $r$ range. (c) Neutron Ti-Ti partial PDF for the two models. Corresponding PDFs for x-ray probe are shown in (d)-(f). Simulations use uniform 0.001 \AA$^2$ ADPs for all atoms, and are scaled to match the data shown in Fig.~\ref{fig:expPDFcomp}.}
\end{figure}

\begin{figure}[tb]
\includegraphics[width=0.485\textwidth]{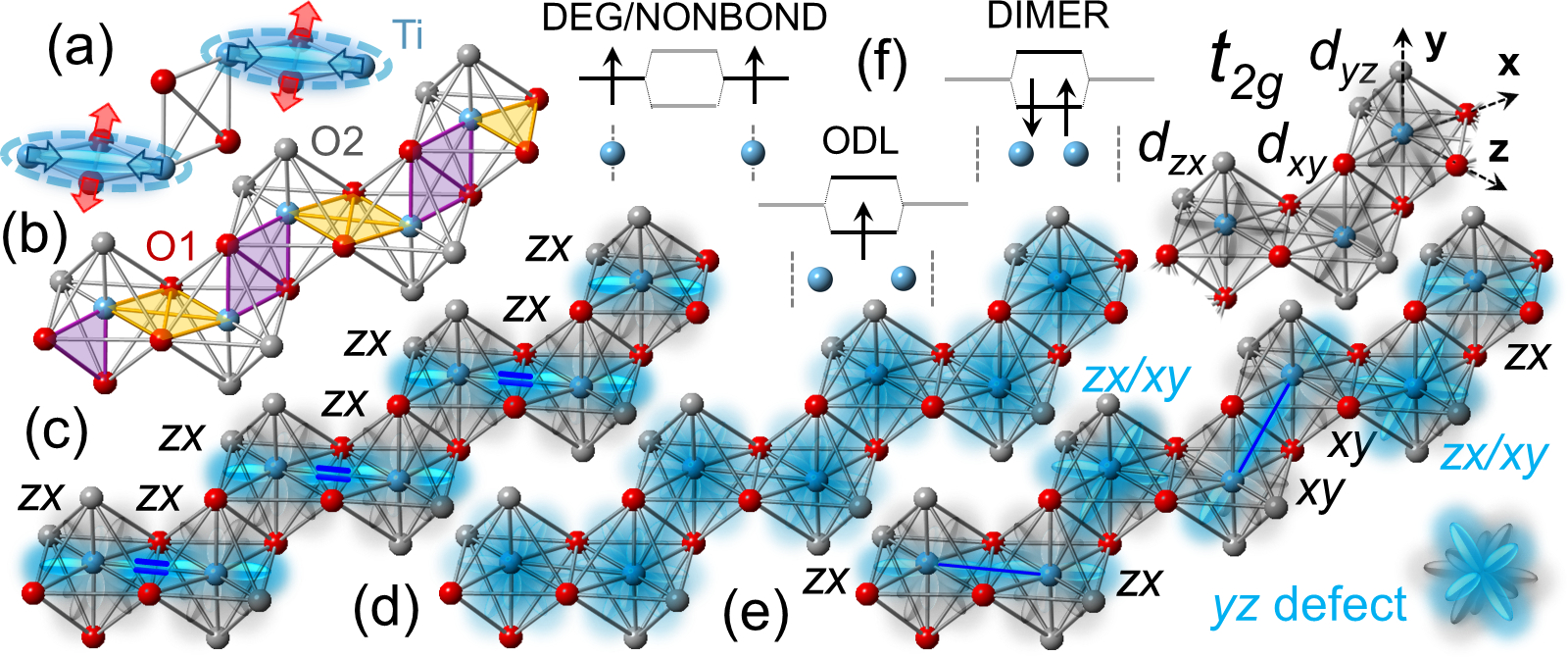}
\caption{\label{fig:ODLfig}\ntso\ orbital considerations. (a) TiO$_{2}$ dimerization plaquettes. (b) The two choices: the zig and the zag. (c) Dimerization of the $zx$ variety within the \p1bar\ structure. (d) Uniform chain with degenerate \t2g\ manifolds as portrayed by the \c2c\ structure model. (e) {\it Local} model of the chain for $T \geq T_{s}$ featuring ODL states with a 6-Ti period. (f) Molecular-orbital (MO) view, counterclockwise, of Ti-Ti contacts with degenerate or non-bonding Ti-Ti contacts, degeneracy-lifted MO, and dimerized Ti-Ti contacts. In the legend, DEG/NONBOND (Ti$^{3+}$), ODL (1e$^{-}$ per Ti-Ti bond), and DIMER (2e$^{-}$ per Ti-Ti bond). Corner insets: orbital geometry of the \t2g\ manifold (upper right), and $yz$ defect discussed in text (lower right).  }
\end{figure}

\ntso\ crystallizes in a monoclinic \c2c\ structure, Fig.~\ref{fig:propfig}(a), featuring characteristic zigzag chains of edge-sharing \tio6\ octahedra, Fig.~\ref{fig:propfig}(b), giving the system a quasi-one-dimensional character~\cite{strel;prb08}.
The chains are embedded in a somewhat disordered SiO$_{4}$ network encompassing Na~\cite{supplementalMaterials}.
Within the chains, the shared-edge O pairs and Ti centers constitute \plaquette\ plaquettes, identical in \c2c, which alternate in orientation, as shown in Fig.~\ref{fig:propfig}(c).
Magnetically active Ti have +3 valence in 3$d^{1}$ configuration ~\cite{redhammer_single-crystal_2003}, confirmed by neutron Rietveld refinement based bond valence sum calculations~\cite{supplementalMaterials}.
\ins{The d}\sout{D}ominant octahedral crystal field splits the Ti 3$d$ orbitals into a partially filled $t_{2g}$ triplet and an empty $e_g$ doublet~\cite{isobe_novel_2002}.
Nominally triply degenerate $t_{2g}$ orbitals~\cite{streltsov_orbital_2017} are oriented toward the \tio6\ edges: {\it xy} and {\it zx} point toward the common edges of the zigzag chains, while {\it yz} is perpendicular to the general chain direction (for illustration see the top right corner inset in Fig.~\ref{fig:ODLfig}).
Partial degeneracy alleviation is expected from slight trigonal distortion of \tio6~\cite{konstantinovic_orbital_2004}, placing the single electron into a two fold degenerate low lying $t_{2g}$-derived ({\it zx}, {\it xy}) doublet~\cite{feiguin_quantum_2019} and rendering the third ({\it yz}) orbital inert~\cite{konstantinovic_orbital_2004,konstantinovic_orbital_2004_ii,silverstein_direct_2014}.
The edge-sharing topology fosters direct ({\it xy, xy}) and ({\it zx, zx}) overlaps of $t_{2g}$ orbitals belonging to neighboring Ti along the chains.
This promotes Ti-Ti dimerization~\cite{konstantinovic_orbital_2004} in the orbitally ordered regime~\cite{strel;prb08} upon cooling below $T_{s}$, lifting the $t_{2g}$ degeneracy and lowering the average symmetry to tricilinic (\p1bar)~\cite{redhammer_single-crystal_2003}.

The average structure change observed in diffraction across the transition is associated with the splitting of Ti-Ti pair distances in the zigzag chains.
The dimerization takes place within the \plaquette\ plaquettes of just one of the two available orientations (zig or zag, Fig.~\ref{fig:ODLfig}(b)).
Consequentially, the Ti-Ti and O-O interatomic distances on the plaquettes bifurcate, Fig.~\ref{fig:propfig}(d): Ti-Ti (3.18~\AA) and O-O (2.74~\AA) contacts on the plaquettes in \c2c\ become (3.11~\AA, 3.22~\AA) and (2.69~\AA, 2.81~\AA) in \p1bar, respectively, with the short Ti-Ti distance (long O-O distance) on a dimerized plaquette~\cite{redhammer_single-crystal_2003} (see Fig.~\ref{fig:ODLfig}(a)).
Neighboring \plaquette\ plaquettes become inequivalent, reflecting Ti dimer and associated bond charge order formation, thus removing the {\it zx/xy} degeneracy.
The \c2c\ and \p1bar\ models explain our neutron Bragg data in the high and low temperature regimes, respectively~\cite{supplementalMaterials}.
All Ti sites participate in dimerization in \p1bar\ but remain equivalent (+3 valence)~\cite{supplementalMaterials,redhammer_single-crystal_2003}.
The fingerprint of the average structural change across $T_{s}$, simulated from crystallographic data above and below the transition~\cite{redhammer_single-crystal_2003} for nPDF (Fig.~\ref{fig:simPDFcomp}(a), (b)) and xPDF (Fig.~\ref{fig:simPDFcomp}(c), (d)), illustrates the expected PDF response should the local structure follow the average behavior.
In Figs.~\ref{fig:simPDFcomp}(c) and~\ref{fig:simPDFcomp}(f) the crystallographically observed Ti-Ti splitting~\cite{redhammer_single-crystal_2003} is shown by scattering-weighted partial PDFs, revealing considerably weaker signal in nPDF than in xPDF case.
Significantly, the pair contributions to PDF of Ti-Ti when compared to O-O are order of magnitude stronger in xPDF case, whereas in nPDF they are 3 times weaker.

\begin{figure}[tb]
\includegraphics[width=0.485\textwidth]{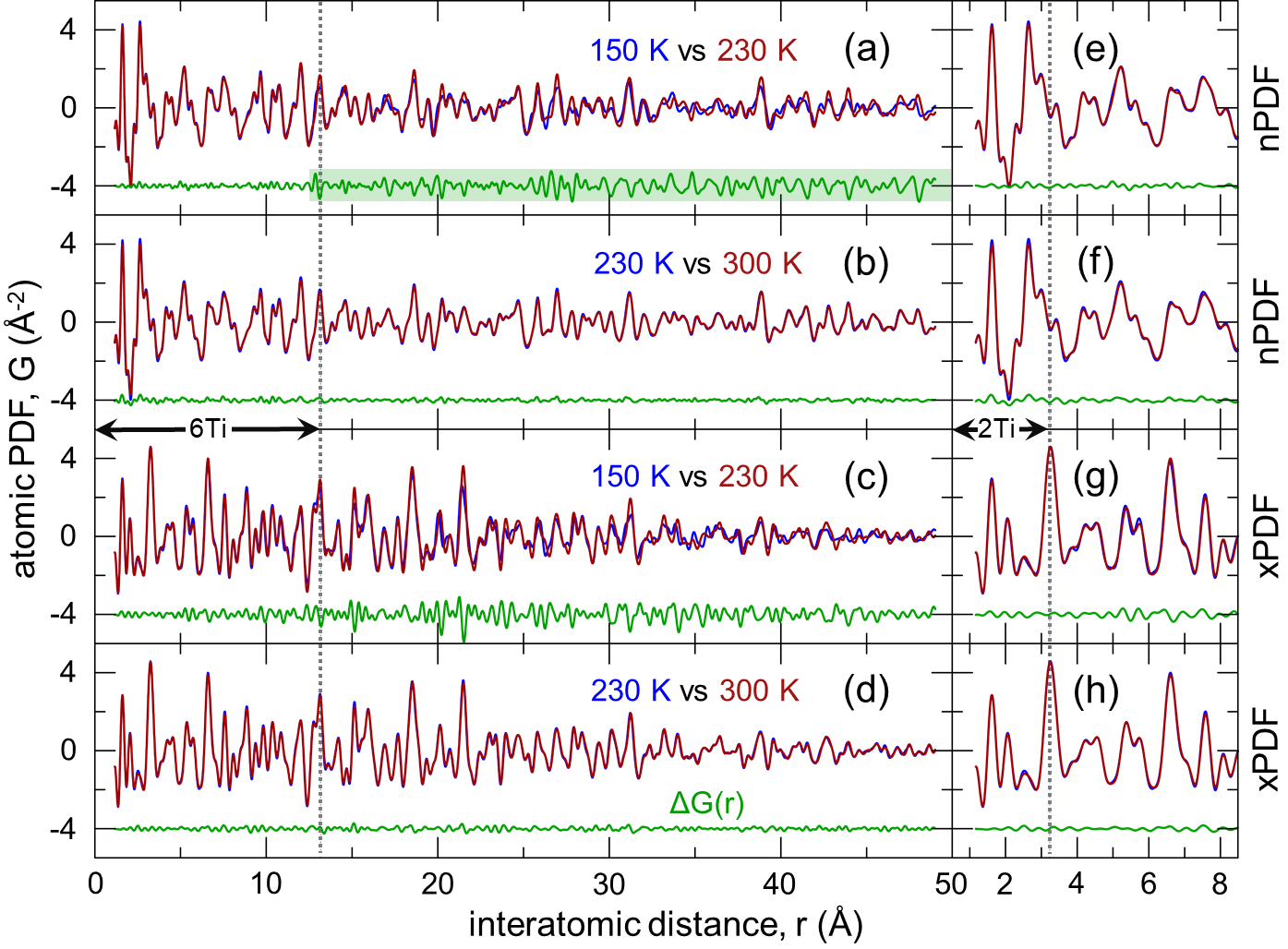}
\caption{\label{fig:expPDFcomp} Comparison of experimental PDFs: Data at temperature below (150~K) and above (230~K) the transition temperature, $T_{s}$, for neutrons (nPDF) over broad (a) and narrow (e) $r$ ranges. Matching X-ray data (xPDF) scaled to nPDF are shown in (c) and (g). Comparison of nPDFs within the same crystallographic phase, \c2c, at 230~K and 300~K, is shown in (b) and (f). The same for xPDF is shown in (d) and (h). Differential PDFs, $\Delta G(r)$ are shown underneath each data, offset for clarity. The vertical dotted lines in panels (a)-(d) and (e)-(h) correspond to the fifth and the first Ti-Ti nearest neighbor distances along the zigzag chains, respectively,  marked also by vertical double arrows as 6~Ti (2~Ti) intrachain interatomic separations.}
\end{figure}

While crystallography may seem to dictate that the lifting of Ti orbital degeneracy and associated dimer formation occur at $T_{s}$, the complexity increases when the local structure information from PDF data is considered.
If we compare the PDF signal from $T$ = 150~K (well below $T_{s}$) to that from $T$ = 230~K (just above $T_{s}$), the difference signal $\Delta G$ for interatomic distances $r>$~15~\AA\ is large and significant, as is to be expected when passing through a structural transition (Fig.~\ref{fig:expPDFcomp}(a) and \ref{fig:expPDFcomp}(c)).
However, and in contrast to the average structure based expectations shown in Fig.~\ref{fig:simPDFcomp}, $\Delta G$ is substantially smaller over the shorter distances ($r <$ 15~\AA) reflecting local structure, as highlighted in Fig.~\ref{fig:expPDFcomp}(e) and ~\ref{fig:expPDFcomp}(g), especially in the nPDF case, which is less sensitive to Ti.

In fact, the local $\Delta G$ observed {\it across} the transition is comparable in magnitude to that observed in a 70~K difference $\Delta G$ which is fully {\it above} the transition (Fig.~\ref{fig:expPDFcomp}(e)-(h)), where only small changes due to thermal motion amplitude variations would be expected.
While the structural transition associated with the dimer formation is clearly apparent in the average structure, the same cannot be said regarding the local structure, revealing a curious local vs average disparity in \ntso.
This may suggest that spin singlet dimers do not disassemble locally on warming across $T_{s}$, in contrast to magnetic susceptibility measurements according to which the spin singlet dimers cannot be retained in the high temperature regime.
We argue below that the transition is not of a trivial order-disorder type, as one may deduce from the nPDF analysis alone~\cite{supplementalMaterials}, Figs.~\ref{fig:expPDFcomp}(a) and \ref{fig:expPDFcomp}(b), but that it has an ODL-type character~\cite{bozin_local_2019} evident from the xPDF data analysis.
In contrast to the differential nPDF signal implying minute change across the transition over the length scale corresponding to $\sim$6 Ti sites, the $\Delta G$ signal in xPDF, Figs.~\ref{fig:expPDFcomp}(c) and \ref{fig:expPDFcomp}(d), suggests that some local structural modification actually does occur at $T_{s}$.

\begin{figure}[tb]
\includegraphics[width=0.485\textwidth]{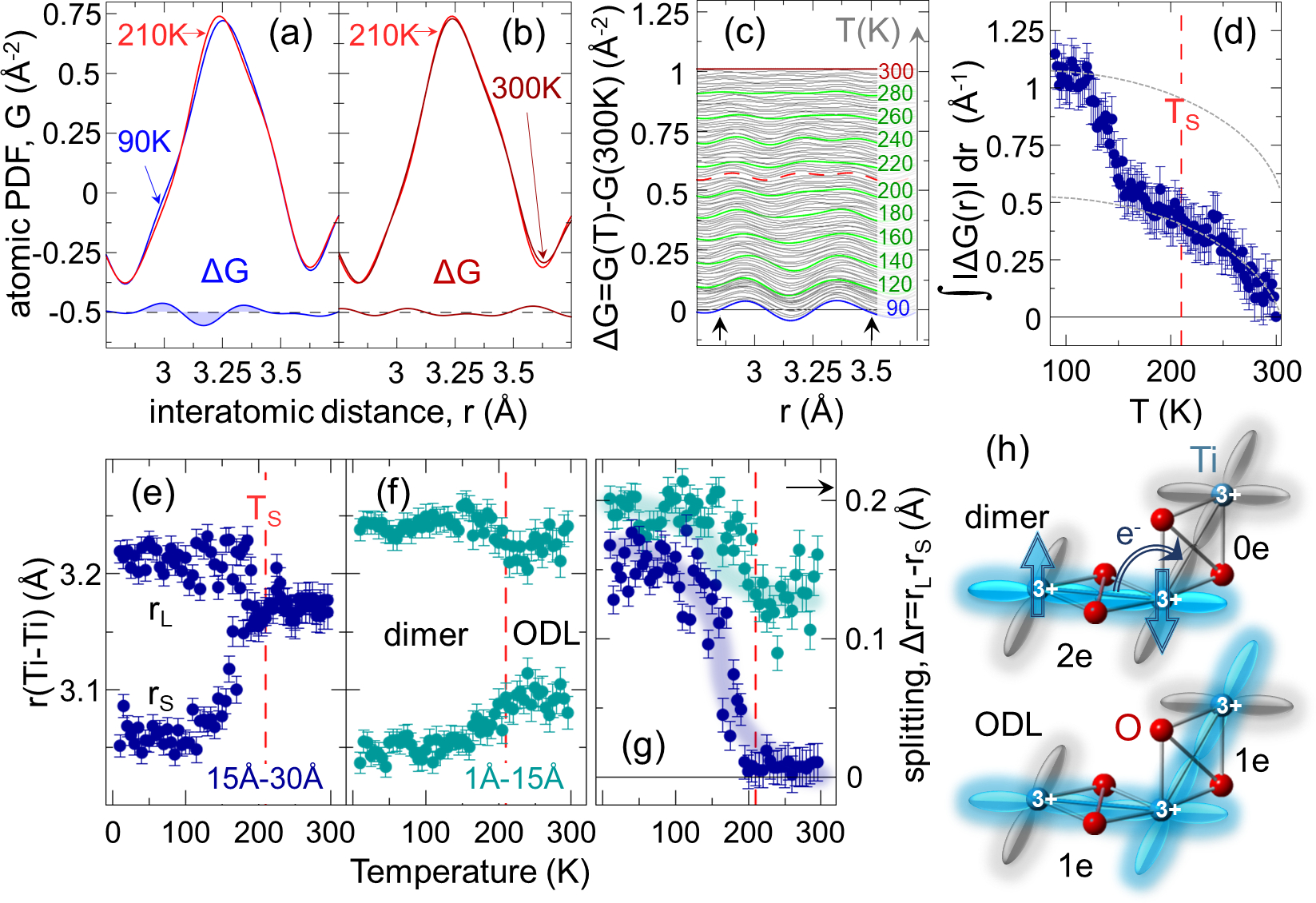}
\caption{\label{fig:DimersDisappear} The spin-singlet dimer disappearance: Comparison of xPDF data at $T_{s}=$ 210~K with (a) 90~K and (b) 300~K data. Differentials  $\Delta G = G(T) - G(T_{s})$ are offset for clarity, revealing the spin-singlet signature (shaded signal) for 90~K set. (c) Waterfall view of the temperature evolution of the xPDF differential using 300~K reference for \rangetxpd\ range ($\Delta $T$ =$ 2~K). Corresponding integrated intensity in the range between vertical black arrows is shown in (d). Dotted gray lines are guides to the eye. Dashed red line in (c) and (d) marks $T_{s}$. The nearest neighbor Ti-Ti distances from \p1bar-based model (see text) fits over (e) \rrangee\ and (f) \rrange\ ranges. Corresponding $r$(Ti-Ti) splittings are shown in (g). (h) The spin-singlet dimer and ODL states sketched as \t2g\ orbital manifold overlaps. The transparency of turquoise color indicates the bond charge filling, as noted.}
\end{figure}

This motivates a closer look at the temperature resolved xPDF data.
Temperature evolution of the PDF differential, $\Delta G(T)$, underneath the Ti-Ti PDF peak at $\sim$3.2~\AA, where the dimer signal should be present, is particularly informative.
Comparing the data at T$=$90~K ($T${$<$}$T_{s}$) and at $T_{s}$ reveals a subtle but clear shift in pair probability from shorter to longer pair distances, Fig.~\ref{fig:DimersDisappear}(a), with an ``M'' shaped feature in $\Delta G(T)$, consistent with removal of the Ti dimer distortion.
When data at $T_{s}$ and at T$=$300~K ({$T$}{$>$}$T_{s}$) are compared, Fig.~\ref{fig:DimersDisappear}(b), the differential is much smaller for a comparable temperature difference.
The nPDF data are not sensitive to this not only due to unfavorable scattering contrast but also because the dimer-related distortions involve Ti and O displacements of opposite sign on the \plaquette\ plaquettes, Fig.~\ref{fig:ODLfig}(a).
For systematic assessment of $\Delta G(T)$ we use 300~K reference for calculating the differentials.
The shift related to the local structure change is inevitably superimposed with thermal broadening effects, which limits this model independent approach close to $T_{s}$.
The evolution with temperature of $\Delta G$ and its integral are shown in Figs.~\ref{fig:DimersDisappear}(c) and  \ref{fig:DimersDisappear}(d), consistent with the dimers disassembling locally at the transition.
The non-thermal contribution to the differential associated with the dimers corresponds to the jump in integral signal seen in Fig.~\ref{fig:DimersDisappear}(d).

To quantify this we fit the temperature-resolved xPDF data using the low temperature \p1bar\ structure for maximum flexibility and over different $r$-ranges.
When the fit range excludes the local structure portion, the optimized model structure refined over 15~\AA $<r<$30~\AA\ range adopts two unique Ti-Ti pair distances below $T_{s}$, and these two distances become degenerate above $T_{s}$, Fig. \ref{fig:DimersDisappear}(e), consistent with dimer elimination.
When the average structure portion of the PDF is excluded and 1~\AA $<r<$15~\AA\ range is used instead, the model structure again adopts two unique Ti-Ti pair distances below $T_{s}$, but these two distances {\it remain distinct} above $T_{s}$, albeit with significantly reduced splitting, Fig. \ref{fig:DimersDisappear}(f).
Thus, at {$T$}{$>$}$T_{s}$ \ntso\ shows a regularization of the Ti chains over long structural length scales, but this regularization is not present locally.
Some degree of degeneracy lifting is apparent above $T_{s}$ observed up to 300~K (Fig. \ref{fig:DimersDisappear}(g)), with splitting of 0.12(4)~\AA\ still present in our 490~K xPDF data.

The behavior where short spin singlet dimer bonds give way to longer local-symmetry-breaking transition metal contacts upon heating above the crystallographic symmetry breaking transition is established as a hallmark of the ODL phenomenology in several spinel dimer systems proximal to \ins{a }localized\ins{-}\sout{ }to\ins{-}\sout{ }itinerant crossover.
Initially observed in \cis~\cite{bozin_local_2019}, and recently also in \mto~\cite{long;arxiv20}, the ODL state is evidenced in their high temperature metallic cubic regimes.
There, at the metal-insulator transition, the spin singlet dimers comprised of pairs of strongly bonded holes (in \cis) or electrons (in \mto) dismount by means of bond charge transfer upon warming, and are succeeded in the metallic phase by twice as many spatiotemporally fluctuating single charge Hund-Mulliken molecular-orbital-like  states~\cite{streltsov_orbital_2017} that lift the \t2g\ degeneracy~\cite{bozin_local_2019,long;arxiv20}.
The dimer and ODL states are shown in Fig.~\ref{fig:ODLfig}(f) using energy diagram representation.
The observed high crystallographic symmetry ensues from three-dimensional spatiotemporal averaging.

In addition to thermal evolution of the transition metal sublattices, the similarity of Ti pyroxenes and the spinels extends to observed pressure effects.
In the spinels, pressure increases \t2g\ orbital overlaps and stabilizes the transition~\cite{furubayashi_pressure_1997,ma_opposite_2017} and the ODL state~\cite{bozin_local_2019}. In \ltso, which is isostructural and isoelectronic to \ntso, $\sim$3.1\% volume reduction pushes $T_{s}$ also to higher temperature~\cite{isobe_novel_2002,supplementalMaterials}, further corroborating the equivalence of the underlying orbital behaviors.
Following the spinel ODL phenomenology, in \ntso\ the Ti dimers exhibit a ``2e-0e''-type bond charge order along the zigzag chains for $T${$<$}$T_{s}$ [Fig.~\ref{fig:ODLfig}(c)], which for {$T$}{$>$}$T_{s}$ converts into the ODL state with the {\it zx/xy} degeneracy being lifted locally, as illustrated in Fig.~\ref{fig:DimersDisappear}(h).

However, pursuing full analogy with the spinels, particularly with \mto\ which has the same Ti \t2g\ filling as in \ntso\ and whose ODL states have a two-orbital (2O-ODL) character~\cite{long;arxiv20} sketched in Fig.~\ref{fig:DimersDisappear}(h), encounters two challenges.
First, in \ntso\ 2O-ODL would imply a single valued Ti-Ti distance distribution akin to the degenerate orbital case portrayed crystallographically, Fig.~\ref{fig:ODLfig}(d). This is {\it not} what is seen experimentally.
The PDF observations impose that, similar to the dimer regime, both shorter (ODL) and longer (non-ODL) Ti-Ti contacts along the zigzag chains are present.
Second, the three-dimensional pyrochlore network of corner-shared transition metal tetrahedra in the spinels provides a basis for reconciling the discrepancy between the average and the local behavior.
Quasi-one-dimensional topology of Ti zigzag chains in \ntso\ necessitates a modified scenario for reconciling the different length scales.

Both challenges can be addressed by considering presence of ``orbital" defects along the chains.
The Ti sites exhibiting either mixed {\it zx/xy} orbital character, such as these depicted in Fig.~\ref{fig:ODLfig}(e), or whose nonbonding {\it yz} orbitals are engaged (bottom right inset to Fig.~\ref{fig:ODLfig}) represent possible defect types.
Notably, local structural correlations extend to $\sim$13.3~\AA, as deduced from the nPDF data comparison in Fig.~\ref{fig:expPDFcomp}(a) and from model assessment of the 290~K xPDF data~\cite{supplementalMaterials} (not shown), and correspond to $\sim$6~Ti distance along the zigzag chain.
This is consistent with a Peierls-like instability at 1/6 filling involving all three Ti orbitals.
Such defects would lead to not only lower-energy excited states than the triplet excitations of the dimerized bonds (see Supplemental Material)~\cite{supplementalMaterials} but also an increased contribution of the entropy term in the free energy at higher
temperature, thus stabilizing the ODL state in \ntso\ which can be considered to have dual ({\it xy,xy})/({\it zx,zx}) character indicated in Fig.~\ref{fig:ODLfig}(e) by solid blue lines connecting affected Ti sites.
In addition to providing entropy driven stabilization, due to lack of direct overlaps with other \t2gs, introduction of the third orbital inevitably implies involvement of intervening oxygen.

In closing, we note that a number of anomalies were observed in the {$T$}{$>$}$T_{s}$ regime of \ntso\ and attributed to various electronic instabilities.
Anomalies include unusual temperature dependence of magnetic susceptibility~\cite{isobe_novel_2002}, the lack of recovery in the muon asymmetry at longer times and lack of sharp change in electronic relaxation rate $\lambda$ at $T_{s}$ in $\mu$SR measurements~\cite{baker_muon-spin_2007}, anomalous and unusually broad phonon modes in Raman~\cite{konstantinovic_orbital_2004,konstantinovic_orbital_2004_ii} and neutron scattering~\cite{silverstein_direct_2014} and infrared reflectivity~\cite{popovic_orbital_2006}, glasslike temperature evolution of thermal conductivity~\cite{rivas-murias_rapidly_2011}, as well as anomalous peak broadening in x-ray diffraction~\cite{silverstein_direct_2014}.
They were assigned to short-range correlations enhancing spin singlet dimer fluctuations~\cite{baker_muon-spin_2007}, orbital disorder~\cite{konstantinovic_orbital_2004,konstantinovic_orbital_2004_ii,popovic_orbital_2006}, rapidly fluctuating orbital occupancy~\cite{rivas-murias_rapidly_2011},
and presence of bond disorder due to orbital fluctuations~\cite{silverstein_direct_2014}, respectively.
Observation of the ODL state, which is presumably dynamic, provides a concrete rationale for their understanding and invites reexamination of the transition mechanism~\cite{hikihara_orbital_2004,wezel_orbital-assisted_2006,silverstein_direct_2014}.
Such high temperature anomalies could in fact be indicators of the ODL state in a diverse class of transition metal systems with active orbital sector~\cite{rivas-murias_rapidly_2011,hiroi_structural_2015,attfield_orbital_2015,katsufuji_universal_2015}, reinforcing the idea of ubiquitous ODL precursor states, extending the phenomenology to strongly correlated electron systems.
%
%
\begin{acknowledgments}
Work at Brookhaven National Laboratory was supported by U.S. Department of Energy, Office of Science, Office of Basic Energy Sciences (DOE-BES) under contract No. DE-SC0012704.
R.S. and H.Z. thank the support from the U.S. Department of Energy under award DE-SC-0020254. Neutron total scattering data were collected at the NOMAD beamline (BL-1B) at the Spallation Neutron Source, a U.S. Department of Energy Office of Science User Facility operated by the Oak Ridge National Laboratory.
X-ray PDF measurements were conducted on beamline 28-ID-1 of the National Synchrotron Light Source II, a U.S. Department of Energy (DOE) Office of Science User Facility operated for the DOE Office of Science by Brookhaven National Laboratory under Contract No. DE-SC0012704.
\end{acknowledgments}


\end{document}